\documentclass[11pt]{article}

\usepackage{amsmath,amsthm,amsfonts,array}

\DeclareMathOperator{\Hom}{Hom}
\DeclareMathOperator{\Com}{Com}

\DeclareMathOperator{\1}{id}
\newcommand{\NN}{\mathbb{N}}
\newcommand{\RR}{\mathbb{R}}
\newcommand{\EEnd}{\mathcal End}
\newcommand{\EE}{\mathcal E}

\renewcommand{\=}{:=}
\renewcommand{\t}{\otimes}

\newcommand{\m}{\overset{\circ}{\mu}}
\newcommand{\A}{{\cal A}}
\newcommand{\pp}{\hat{p}}
\newcommand{\ppp}{p_0}
\newcommand{\q}{\hat{q}}
\newcommand{\muu}{\hat{\mu}}
\newcommand{\qxi}{\hat{\xi}}
\newcommand{\Q}{\hat{Q}}
\newcommand{\PP}{\hat{P}}
\newcommand{\ee}{\varepsilon}
\renewcommand{\H}{\hat{H}}
\newcommand{\e}{\hat{\varepsilon}}
\newtheorem{thm}{Theorem}[section]

 \newtheorem{lemma}[thm]{Lemma}
 \newtheorem{cor}[thm]{Corollary}
 
\theoremstyle{definition}
 \newtheorem{defn}[thm]{Definition}

\theoremstyle{definition}

\theoremstyle{definition}
 \newtheorem{rem}[thm]{Remark}

\numberwithin{equation}{section}
\numberwithin{table}{section}

\begin{document}

\title{\bf\LARGE Quantum counterparts of \\
VII$_{a}$, III$_{a=1}$, VI$_{a\neq1}$ over harmonic oscillator \\
in semiclassical approximation}
\author{\Large Eugen Paal and J\"uri Virkepu}
\date{}
\maketitle
\thispagestyle{empty}

\begin{abstract}
Operadic Lax representations for the harmonic oscillator are used to construct the quantum counterparts of some real three dimensional Lie algebras.  The Jacobi operators of these quantum algebras are studied in semiclassical approximation.
\end{abstract}

\section{Introduction and outline of the paper}

In Hamiltonian formalism, a mechanical system is described by the
canonical variables $q^i,p_i$ and their time evolution is prescribed
by the Hamiltonian equations
\begin{equation}
\label{ham} \dfrac{dq^i}{dt}=\dfrac{\partial H}{\partial p_i}, \quad
\dfrac{dp_i}{dt}=-\dfrac{\partial H}{\partial q^i}
\end{equation}
By a Lax representation \cite{Lax68} of a mechanical system
one means such a pair $(L,M)$ of matrices (linear operators) $L,M$
that the above Hamiltonian system may be represented as the Lax
equation
\begin{equation}
\label{lax} \dfrac{dL}{dt}= ML-LM
\end{equation}
Thus, from the algebraic point of view, mechanical systems may be
represented by linear operators, i.e by  linear maps $V\to V$  of a
vector space $V$. As a generalization of this one can pose the
following question \cite{Paal07}: how to describe the time evolution
of the linear operations (multiplications) $V^{\t n}\to V$?

The algebraic operations (multiplications) can be seen as an example
of the \emph{operadic} variables \cite{Ger}. If an operadic system
depends on time one can speak about \emph{operadic dynamics}
\cite{Paal07}. The latter may be introduced by simple and natural
analogy with the Hamiltonian dynamics. In particular, the time
evolution of the operadic variables may be given by the operadic Lax
equation. In \cite{PV07,PV08,PV08-1}, the low-dimensional binary operadic Lax representations for the harmonic oscillator were constructed.
In \cite{PV08-2} it was shown how the operadic Lax representations are related to the conservation of energy.

In this paper, the operadic Lax representations for the harmonic oscillator are used to construct the quantum counterparts of some real three dimensional Lie algebras.  The Jacobi operators of these quantum algebras are studied in semiclassical approximation.

\section{Endomorphism operad and Gerstenhaber brackets}

Let $K$ be a unital associative commutative ring, $V$ be a unital
$K$-module, and $\EE_V^n\= {\EEnd}_V^n\= \Hom(V^{\t n},V)$
($n\in\NN$). For an \emph{operation} $f\in\EE^n_V$, we refer to $n$
as the \emph{degree} of $f$ and often write (when it does not cause
confusion) $f$ instead of $\deg f$. For example, $(-1)^f\= (-1)^n$,
$\EE^f_V\=\EE^n_V$ and $\circ_f\= \circ_n$. Also, it is convenient
to use the \emph{reduced} degree $|f|\= n-1$. Throughout this paper,
we assume that $\t\= \t_K$.

\begin{defn}[endomorphism operad \cite{Ger}]
\label{HG} For $f\t g\in\EE_V^f\t\EE_V^g$ define the \emph{partial
compositions}
\[
f\circ_i g\= (-1)^{i|g|}f\circ(\1_V^{\t i}\t g\t\1_V^{\t(|f|-i)})
\quad \in\EE^{f+|g|}_V,
         \quad 0\leq i\leq |f|
\]
The sequence $\EE_V\= \{\EE_V^n\}_{n\in\NN}$, equipped with the
partial compositions $\circ_i$, is called the \emph{endomorphism
operad} of $V$.
\end{defn}

\begin{defn}[total composition \cite{Ger}]
The \emph{total composition}
$\circ \:\EE^f_V\t\EE^g_V\to\EE^{f+|g|}_V$ is defined by
\[
f\circ g\= \sum_{i=0}^{|f|}f\circ_i g\quad \in \EE_V^{f+|g|}, \quad |\circ|=0
\]
The pair $\Com\EE_V\= \{\EE_V,\circ\}$ is called the \emph{composition
algebra} of $\EE_V$.
\end{defn}

\begin{defn}[Gerstenhaber brackets \cite{Ger}]
The  \emph{Gerstenhaber brackets} $[\cdot,\cdot]$ are defined in
$\Com\EE_V$ as a graded commutator by
\[
[f,g]\= f\circ g-(-1)^{|f||g|}g\circ f=-(-1)^{|f||g|}[g,f],\quad
|[\cdot,\cdot]|=0
\]
\end{defn}

The \emph{commutator algebra} of $\Com \EE_V$ is denoted as
$\Com^{-}\!\EE_V\= \{\EE_V,[\cdot,\cdot]\}$. One can prove (e.g
\cite{Ger}) that $\Com^-\!\EE_V$ is a \emph{graded Lie algebra}. The
Jacobi identity reads
\[
(-1)^{|f||h|}[f,[g,h]]+(-1)^{|g||f|}[g,[h,f]]+(-1)^{|h||g|}[h,[f,g]]=0
\]

\section{Operadic dynamics and Lax equation}

Assume that $K\= \RR$ or $K\= \mathbb{C}$ and operations are
differentiable. Dynamics in operadic systems (operadic dynamics) may
be introduced by

\begin{defn}[operadic Lax pair \cite{Paal07}]
Allow a classical dynamical system to be described by the
Hamiltonian system \eqref{ham}. An \emph{operadic Lax pair} is a
pair $(\mu,M)$ of homogeneous operations $\mu,M\in\EE_V$, such that the
Hamiltonian system  (\ref{ham}) may be represented as the
\emph{operadic Lax equation}
\[
\frac{d\mu}{dt}=[M,\mu]\= M\circ\mu-(-1)^{|M||\mu|}\mu\circ M
\]
The pair $(L,M)$ is also called an \emph{operadic Lax representations} of/for Hamiltonian system \eqref{ham}.
Evidently, the degree constraints $|M|=|L|=0$ give rise to ordinary
Lax equation (\ref{lax}) \cite{Lax68}. In this paper we assume that $|M|=0$.
\end{defn}

The Hamiltonian of the harmonic oscillator (HO) is
\[
H(q,p)=\frac{1}{2}(p^2+\omega^2 q^2)
\]
Thus, the Hamiltonian system of HO reads
\begin{equation}
\label{eq:h-osc} \frac{dq}{dt}=\frac{\partial H}{\partial p}=p,\quad
\frac{dp}{dt}=-\frac{\partial H}{\partial q}=-\omega^2q
\end{equation}
If $\mu$ is a linear algebraic operation we can use the above
Hamilton equations to obtain
\[
\dfrac{d\mu}{dt} =\dfrac{\partial\mu}{\partial
q}\dfrac{dq}{dt}+\dfrac{\partial\mu}{\partial p}\dfrac{dp}{dt}
=p\dfrac{\partial\mu}{\partial
q}-\omega^2q\dfrac{\partial\mu}{\partial p}
 =[M,\mu]
\]
Therefore, we get the following linear partial differential equation
for $\mu(q,p)$:
\begin{equation}
\label{eq:diff}
p\dfrac{\partial\mu}{\partial
q}-\omega^2q\dfrac{\partial\mu}{\partial p}=[M,\mu]
\end{equation}
By integrating \eqref{eq:diff} one can get collections of operations called \cite{Paal07} the \emph{operadic} (Lax representations for/of) harmonic oscillator. 

\section{3D binary anti-commutative operadic Lax representations for harmonic oscillator}

\begin{lemma}
\label{lemma:harmonic3} Matrices
\[
L\=\begin{pmatrix}
    p & \omega q & 0 \\
    \omega q & -p & 0 \\
    0 & 0 & 1 \\
  \end{pmatrix},\quad
M\=\frac{\omega}{2}
\begin{pmatrix}
    0 & -1 &0\\
1 & 0 & 0\\
0 & 0 & 0
  \end{pmatrix}
\]
give a 3-dimensional Lax representation for the harmonic oscillator.
\end{lemma}

\begin{defn}[quasi-canonical coordinates]
For the HO, define its  \emph{quasi-canonical coordinates} $Q$ and $P$ by
\begin{equation}
\label{eq:def_A}
P^{2} - Q^{2}=2p,\quad
QP=\omega q
\end{equation}
\end{defn}

\begin{rem}
Note that these constraints easily imply
\[
P^2+Q^2=2\sqrt{2H}
\]
\end{rem}

\begin{thm}[\cite{PV08-1}]
\label{thm:main}
Let $C_{\nu}\in\mathbb{R}$ ($\nu=1,\ldots,9$) be
arbitrary real--valued parameters, such that
\begin{equation}
\label{eq:cond} C_2^2+C_3^2+C_5^2+C_6^2+C_7^2+C_8^2\neq0
\end{equation}
Let $M$ be defined as in Lemma \ref{lemma:harmonic3} and 
$\mu: V\otimes V\to V$ be a binary operation in a 3 dimensional real vector space $V$ with the coordinates
\begin{equation}\label{eq:theorem}
\begin{cases}
\mu_{11}^{1}=\mu_{22}^{1}=\mu_{33}^{1}=\mu_{11}^{2}=\mu_{22}^{2}=\mu_{33}^{2}=\mu_{11}^{3}=\mu_{22}^{3}=\mu_{33}^{3}=0\\
\mu_{23}^{1}=-\mu_{32}^{1}=C_2p-C_3\omega q-C_4\\
\mu_{13}^{2}=-\mu_{31}^{2}=C_2p-C_3\omega q+C_4\\
\mu_{31}^{1}=-\mu_{13}^{1}=C_2\omega q+C_3p-C_1\\
\mu_{23}^{2}=-\mu_{32}^{2}=C_2\omega q+C_3p+C_1\\
\mu_{12}^{1}=-\mu_{21}^{1}=C_5 P + C_6 Q\\
\mu_{12}^{2}=-\mu_{21}^{2}=C_5 Q - C_6 P\\
\mu_{13}^{3}=-\mu_{31}^{3}=C_7 P + C_8 Q\\
\mu_{23}^{3}=-\mu_{32}^{3}=C_7 Q - C_8 P\\
\mu_{12}^{3}=-\mu_{21}^{3}=C_9
\end{cases}
\end{equation}
Then $(\mu,M)$ is an operadic Lax pair for HO.
\end{thm}

\section{Initial conditions}

Specify the coefficients $C_{\nu}$ in Theorem \ref{thm:main} by the
initial conditions
\[
\left. \mu\right|_{t=0}=\m{}_,\quad
\left.p\right|_{t=0}
=p_0,\quad \left. q\right|_{t=0}=0
\]
Denoting $E\=H|_{t=0}$, the latter together with \eqref{eq:def_A} yield the initial
conditions for $Q$ and $P$:
\[
\begin{cases}
\left.\left(P^{2}+Q^{2}\right)\right|_{t=0}=2\sqrt{2E}\\
\left.\left(P^{2}-Q^{2}\right)\right|_{t=0}=2p_0\\
\left.PQ\right|_{t=0}=0
\end{cases}
\quad \Longleftrightarrow \quad
\begin{cases}
\ppp\!\!>0\\
\left.P^{2}\right|_{t=0}=2p_0\\
\left.Q\right|_{t=0}=0
\end{cases}
\vee\quad
\begin{cases}
\ppp<0\\
\left.P\right|_{t=0}=0\\
\left.Q^2\right|_{t=0}=-2p_0
\end{cases}
\]
In what follows assume that $p_0>0$ and $P|_{t=0}=\sqrt{2p_0}$. The other cases
can be treated similarly. Note that in this case $p_0=\sqrt{2E}$. From \eqref{eq:theorem} we get the following linear system:
\begin{equation}
\label{eq:constants} 
\left\{
  \begin{array}{lll}
C_1=\frac{1}{2}\left(\overset{\circ}{\mu}{}_{23}^{2}-\overset{\circ}{\mu}{}_{31}^{1}\right),&
C_2=\frac{1}{2\ppp}\left(\overset{\circ}{\mu}{}_{13}^{2}+\overset{\circ}{\mu}{}_{23}^{1}\right),&
C_3=\frac{1}{2\ppp}\left(\overset{\circ}{\mu}{}_{23}^{2}+\overset{\circ}{\mu}{}_{31}^{1}\right)\vspace{1mm}\\
C_4=\frac{1}{2}\left(\overset{\circ}{\mu}{}_{13}^{2}-\overset{\circ}{\mu}{}_{23}^{1}\right),&
C_5=\frac{1}{\sqrt{2\ppp}}\overset{\circ}{\mu}{}_{12}^{1},&
C_6=-\frac{1}{\sqrt{2\ppp}}\overset{\circ}{\mu}{}_{12}^{2}\vspace{1mm}\\
C_7=\frac{1}{\sqrt{2\ppp}}\overset{\circ}{\mu}{}_{13}^{3},&
C_8=-\frac{1}{\sqrt{2\ppp}}\overset{\circ}{\mu}{}_{23}^{3},&
C_9=\overset{\circ}{\mu}{}_{12}^{3}
\end{array}
\right.
\end{equation}

\section{VII$_{a}$, III$_{a=1}$, VI$_{a\neq1}$}

We study only the algebras VII$_{a}$, III$_{a=1}$, VI$_{a\neq1}$ from the Bianchi classification of the real three dimensional Lie algebras \cite{Landau80}.
The structure equations of the 3-dimensional real Lie algebras can be presented
as follows:
\[
[e_1,e_2]=-\alpha e_2+n^{3}e_3,\quad
[e_2,e_3]=n^{1}e_1,\quad
[e_3,e_1]=n^{2}e_2+\alpha e_3
\]
The values of the parameters $\alpha,n^{1}, n^{2},n^{3}$  and the corresponding structure constants for II, VII$_{a}$, III$_{a=1}$, VI$_{a\neq1}$  are presented in Table \ref{table:Bianchi1}. Note that II is the real three dimensional Heisenberg algebra.
\begin{table}[ht]
\begin{center}
\begin{tabular}{|c||c||c|c|c||c|c|c|c|c|c|c|c|c|c|c|}\hline
Bianchi type & $\alpha$ & $n^{1}$ & $n^{2}$ & $n^{3}$ &
$\overset{\circ}{\mu}{}_{12}^{1}$ &
$\overset{\circ}{\mu}{}_{12}^{2}$ &
$\overset{\circ}{\mu}{}_{12}^{3}$ &
 $\overset{\circ}{\mu}{}_{23}^{1}$ & $\overset{\circ}{\mu}{}_{23}^{2}$ & $\overset{\circ}{\mu}{}_{23}^{3}$
  & $\overset{\circ}{\mu}{}_{31}^{1}$ & $\overset{\circ}{\mu}{}_{31}^{2}$ &
  $\overset{\circ}{\mu}{}_{31}^{3}$\\\hline\hline
VII$_{a}$& $a$ & 0 & $1$ & $1$ & 0 & $-a$ & $1$ & 0 & 0 & 0 & 0 &
$1$ & $a$ \\\hline
III$_{a=1}$& 1 & 0 & $1$ & $-1$ & 0 & $-1$ & $-1$ & 0 & 0 & 0 & 0 &
$1$ & $1$
\\\hline
VI$_{a\neq 1}$& $a$ & 0 & $1$ & $-1$ & 0 & $-a$ & $-1$ & 0 & 0 & 0 &
0 & $1$ & $a$
\\\hline
\end{tabular}
\end{center}
\caption{VII$_{a}$, III$_{a=1}$, VI$_{a\neq1}$. Here $a>0$.}
\label{table:Bianchi1}
\end{table}

\section{VII$_{a}^{t}$, III$_{a=1}^{t}$, VI$_{a\neq1}^{t}$}

By using the structure constants of the 3-dimensional Lie algebras
in the Bianchi classification, Theorem \ref{thm:main} and relations
\eqref{eq:constants} one can propose that evolution of VII$_{a}$, III$_{a=1}$, VI$_{a\neq1}$ can be prescribed \cite{PV08-2}
as given in Table \ref{table:Bianchi3}.

\begin{table}[ht]
\begin{center}\setlength\extrarowheight{4pt}
\begin{tabular}{|c||c|c|c|c|c|c|c|c|c|c|c|}\hline
Dynamical Bianchi type & $\mu_{12}^{1}$ & $\mu_{12}^{2}$ &
$\mu_{12}^{3}$ & $\mu_{23}^{1}$ & $\mu_{23}^{2}$ & $\mu_{23}^{3}$ &
$\mu_{31}^{1}$ & $\mu_{31}^{2}$ &  $\mu_{31}^{3}$
\\[1.5ex]\hline\hline
VII$^{t}_a$ & $\frac{aQ}{\sqrt{2p_0}}$ &
$\frac{-aP}{\sqrt{2p_0}}$ & $1$ & $\frac{p-p_0}{-2p_0}$ &
$\frac{\omega q}{-2p_0}$ & $\frac{-aQ}{\sqrt{2p_0}}$ &
$\frac{\omega q}{-2p_0}$ & $\frac{p+p_0}{2p_0}$ &
$\frac{aP}{\sqrt{2p_0}}$
\\ [1.5ex] \hline
III$_{a=1}^{t}$ & $\frac{Q}{\sqrt{2p_0}}$ &
$\frac{-P}{\sqrt{2p_0}}$ & $-1$ & $\frac{p-p_0}{-2p_0}$ &
$\frac{\omega q}{-2p_0}$ & $\frac{-Q}{\sqrt{2p_0}}$ & $\frac{\omega
q}{-2p_0}$ & $\frac{p+p_0}{2p_0}$ & $\frac{P}{\sqrt{2p_0}}$
\\ [1.5ex] \hline
VI$_{a\neq1}^{t}$ & $\frac{aQ}{\sqrt{2p_0}}$ &
$\frac{-aP}{\sqrt{2p_0}}$ & $-1$ & $\frac{p-\ppp}{-2p_0}$ &
$\frac{\omega q}{-2p_0}$ & $\frac{-aQ}{\sqrt{2p_0}}$ &
$\frac{\omega q}{-2p_0}$ & $\frac{p+p_0}{2p_0}$ &
$\frac{aP}{\sqrt{2p_0}}$
\\ [1.5ex] \hline
\end{tabular}
\end{center}
\caption{VII$_{a}^{t}$, III$_{a=1}^{t}$, VI$_{a\neq1}^{t}$. Here $p_0=\sqrt{2E}$.}
\label{table:Bianchi3}
\end{table}

\section{VII$_{a}^{\hbar}$, III$_{a=1}^{\hbar}$, VI$_{a\neq1}^{\hbar}$ and quantum Jacobi operators}

By using the algebras VII$_{a}^{t}$, III$_{a=1}^{t}$, VI$_{a\neq1}^{t}$
from Table \ref{table:Bianchi3}, one can propose \cite{PV09-1} their quantum counterparts 
VII$_{a}^{\hbar}$, III$_{a=1}^{\hbar}$, VI$_{a\neq1}^{\hbar}$ as follows.

Let $\A_{HO}$ denote the state space of the quantum harmonic oscillator and $\{e_1,e_2,\ldots\}$ be its basis.
By using  Table \ref{table:Bianchi4} we define the structure equations in $\A_{HO}$ by
\[
[e_i,e_j]_\hbar:=\muu_{ij}^{s} e_s
\]
where the structure operators $\muu_{ij}^{s}$ for $i,j,s\leq3$ are defined by Table \ref{table:Bianchi4} and $\muu_{ij}^{s}:=0$ for $i,j,s>3$. 
\begin{table}[ht]
\begin{center}\setlength\extrarowheight{4pt}
\begin{tabular}{|c||c|c|c|c|c|c|c|c|c|c|c|}\hline
Quantum Bianchi type & $\muu_{12}^{1}$ & $\muu_{12}^{2}$ &
$\muu_{12}^{3}$ & $\muu_{23}^{1}$ & $\muu_{23}^{2}$ &
$\muu_{23}^{3}$ & $\muu_{31}^{1}$ & $\muu_{31}^{2}$ &
$\muu_{31}^{3}$
\\[1.5ex]\hline\hline
VII$^{\hbar}_a$ & $\frac{a\Q}{\sqrt{2\ppp}}$ &
$\frac{-a\PP}{\sqrt{2\ppp}}$ & $1$ & $\frac{\pp-\ppp}{-2\ppp}$ &
$\frac{\omega \q}{-2\ppp}$ & $\frac{-a\Q}{\sqrt{2\ppp}}$ &
$\frac{\omega \q}{-2\ppp}$ & $\frac{\pp+\ppp}{2\ppp}$ &
$\frac{a\PP}{\sqrt{2\ppp}}$
\\ [1.5ex] \hline
III$_{a=1}^{\hbar}$ & $\frac{\Q}{\sqrt{2\ppp}}$ &
$\frac{-\PP}{\sqrt{2\ppp}}$ & $-1$ & $\frac{\pp-\ppp}{-2\ppp}$ &
$\frac{\omega \q}{-2\ppp}$ & $\frac{-\Q}{\sqrt{2\ppp}}$ & $\frac{\omega
\q}{-2\ppp}$ & $\frac{\pp+\ppp}{2\ppp}$ & $\frac{\PP}{\sqrt{2\ppp}}$
\\ [1.5ex] \hline
VI$_{a\neq1}^{\hbar}$ & $\frac{a\Q}{\sqrt{2\ppp}}$ &
$\frac{-a\PP}{\sqrt{2\ppp}}$ & $-1$ & $\frac{\pp-\ppp}{-2\ppp}$ &
$\frac{\omega \q}{-2\ppp}$ & $\frac{-a\Q}{\sqrt{2\ppp}}$ &
$\frac{\omega \q}{-2\ppp}$ & $\frac{\pp+\ppp}{2\ppp}$ &
$\frac{a\PP}{\sqrt{2\ppp}}$
\\ [1.5ex] \hline
\end{tabular}
\end{center}
\caption{VII$_{a}^{\hbar}$, III$_{a=1}^{\hbar}$, VI$_{a\neq1}^{\hbar}$.}
\label{table:Bianchi4}
\end{table}
For $x,y\in \A_{HO}$, their quantum multiplication is defined by
\[
[x,y]_\hbar
:=\hat{\mu}^{i}_{jk} x^{j}y^{k} e_i
=\hat{\mu}^{1}_{jk} x^{j}y^{k} e_1 
+\hat{\mu}^{2}_{jk} x^{j}y^{k} e_2
+\hat{\mu}^{3}_{jk} x^{j}y^{k} e_3
\]
where we omitted the trivial terms, because 
$\hat{\mu}^{i}_{jk}=0$ for $i>3$. Then the quantum Jacobi operator is defined by
\begin{align*}
\hat{J}_\hbar(x;y;z)
&:=[x,[y,z]_\hbar]_\hbar+[y,[z,x]_\hbar]_\hbar+[z,[x,y]_\hbar]_\hbar\\
&\,\,=\hat{J}^1_\hbar(x;y;z)e_1+\hat{J}^2_\hbar(x;y;z)e_2+\hat{J}^3_\hbar(x;y;z)e_3
\end{align*}
where we again omitted the trivial terms, because $\hat{J}^i_\hbar=0$ for $i>3$. In \cite{PV09-1} the quantum Jacobi operators were calculated for all real three dimensional Lie algebras. Here we concentrate only on III$_{a=1}^{\,\hbar}$, VI$_{a\neq1}^{\,\hbar}$, and VII$^{\,\hbar}_a$. 
Denote
\begin{equation*}
(x,y,z)\=
\begin{vmatrix}
 x^{1} & x^{2} & x^{3} \\
 y^{1} & y^{2} & y^{3} \\
z^{1} & z^{2} & z^{3} \\
\end{vmatrix},\quad
\qxi^{1}\=\omega\q\Q + (\pp - \ppp)\PP, \quad
\qxi^{2}\=\omega\q\PP - (\pp + \ppp)\Q
\end{equation*}
Recall
\begin{thm}[\cite{PV09-1}]
The Jacobi operator components of 
VI$_{a\neq1}^{\,\hbar}$ and VII$^{\,\hbar}_a$ read
\[
      \hat{J}^{1}_\hbar(x;y;z)=-\frac{a(x,y,z)}{\sqrt{2\ppp^{3}}}\qxi^{1},\quad
      \hat{J}^{2}_\hbar(x;y;z)=-\frac{a(x,y,z)}{\sqrt{2\ppp^{3}}}\qxi^{2},\quad
      \hat{J}^{3}_\hbar(x;y;z)=\frac{a^{2}(x,y,z)}{\ppp}[\PP,\Q]
 \]
For III$_{a=1}^{\,\hbar}$ one has the same formulae with $a=1$.
\end{thm}

\section{Semiclassical quantum conditions}
\label{a1}

\begin{thm}[Poisson brackets of quasi-canonical coordinates]
\label{eq:quasi_poisson}
The quasi-canonical coordinates $Q$ and $P$ satisfy the relations
\begin{equation}
\label{eq:poisson_A}
\{P,P\}=0=\{Q,Q\},\quad \{P,Q\}=\ee\=\frac{\omega}{2\sqrt{2H}}
\end{equation}
\end{thm}

\begin{proof}
While the first two relations in \eqref{eq:poisson_A} are evident, we have only to check the third one. Using several times the Leibniz rule for the Poisson brackets, calculate:
\begin{align*}
2\omega=2\omega \{p,q\}
&=\{P^2-Q^2, PQ\}\\
&=\{P^2, PQ\}-\{Q^2, PQ\}\\
&=P\{P^2, Q\}-\{Q^2,P\}Q\\
&=P\{PP, Q\}-\{QQ,P\}Q\\
&=2(P^2+Q^2)\{P,Q\}\\
&=4\sqrt{2H}\{P,Q\}
\tag*{\qed}
\end{align*}
\renewcommand{\qed}{}
\end{proof}

When performing the quantization of the quasi-canonical variables, 
 we shall use the Schr\"o\-dinger picture, i.e the operators $\q,\pp,\H$ and $\Q,\PP$ do not depend on time.
Denote by $[\cdot,\cdot]$ the ordinary commutator bracketing.
Following the canonical quantization prescription, the quasi-canonical coordinates would satisfy in the semiclassical limit ($\hbar\to 0$) the constraints
\begin{equation}
\label{eq:def_qA}
\PP^2+\Q^2\approx 2\sqrt{2\H},\quad
\PP^2-\Q^2\approx 2\pp,\quad
\PP\Q+\Q\PP\approx  2\omega \q
\end{equation}
and the  \emph{quasi-canonical commutation relations} (quasi-CCR)  read as follows:
\begin{equation}
\label{eq:qpoisson_A}
[\PP,\PP]=0=[\Q,\Q],\quad 
[\PP,\Q]\approx \frac{\hbar}{i}\e \=\frac{\hbar}{i}\frac{\omega}{2\sqrt{2\hat{H}}}
\end{equation}

\section{Semiclassical approximation of the Jacobi operator}
\label{a2}

\begin{thm}
Let constraints \eqref{eq:def_qA} and \eqref{eq:qpoisson_A} hold.
Then we have:
\begin{align*}
\hat{J}^{1}_\hbar(x;y;z)
&\approx \frac{a(x,y,z)}{\sqrt{2p_0^3}} 
\left[\PP\left(\sqrt{2E} - \sqrt{2\H}\right)
 - \frac{\hbar}{i}\Q \frac{\e}{2}  \right]\\
\hat{J}^{2}_\hbar(x;y;z)
&\approx  \frac{a(x,y,z)}{\sqrt{2p_0^3}} 
\left[\Q\left(\sqrt{2E} - \sqrt{2\H}\right)
+ \frac{\hbar}{i}\PP \frac{\e}{2} \right]\\
\hat{J}^{3}_\hbar(x;y;z)
&\approx \frac{\hbar}{i}  \frac{a^{2}(x,y,z)}{p_0} \e
\tag*{\qed}
\end{align*}
\end{thm}

\begin{proof}
Using relations \eqref{eq:def_qA} and \eqref{eq:qpoisson_A} first calculate:
\begin{align*}
\qxi^{1}
&\= \omega\q\Q + (\pp - p_0)\PP\\
&\approx \frac{1}{2}(\PP\Q+\Q\PP)\Q +\frac{1}{2}(\PP^{2}-\Q^{2})\PP - p_0 \PP\\
&\approx \frac{1}{2}(\PP\Q^{2}+\Q\PP\Q +\PP^{3} - \Q^{2}\PP)  -p_0\PP\\
&\approx \frac{1}{2}\left[\Q(\PP\Q - \Q\PP)+\PP(\Q^{2} + \PP^{2})\right]-p_0\PP\\
&\approx \frac{1}{2}\Q[\PP,\Q]+\frac{1}{2}\PP(\PP^{2} + \PP^{2})  -p_0\PP\\
&\approx \frac{\hbar}{i}\Q\frac{\e}{2} +\PP\sqrt{2\H} - \sqrt{2E}\PP\\
&\approx \frac{\hbar}{i}\Q\frac{\e}{2} +\PP\left(\sqrt{2\H} - \sqrt{2E}\right)
 \end{align*}
Next calculate
\begin{align*}
\qxi^{2}
&\= \omega\q\PP- (\pp + p_0)\Q\\
&\approx \frac{1}{2}(\PP\Q+\Q\PP)\PP-\frac{1}{2}(\PP^{2}-\Q^{2})\Q -p_0 \Q\\
&\approx \frac{1}{2}(\PP\Q\PP+\Q\PP^{2} -\PP^{2}\Q+ \Q^{3})  -p_0\Q\\
&\approx \frac{1}{2}\left[\PP(\Q\PP - \PP\Q)+\Q(\PP^{2} + \Q^{2})\right]-p_0\Q\\
&\approx -\frac{1}{2}\PP[\PP,\Q]+\frac{1}{2}\Q(\PP^{2} + \Q^{2})  -p_0\Q\\
&\approx -\frac{\hbar}{i}\PP\frac{\e}{2} +\Q\sqrt{2\H} - \sqrt{2E} \Q\\
&\approx -\frac{\hbar}{i}\PP\frac{\e}{2} +\Q\left(\sqrt{2\H} - \sqrt{2E}\right)
\tag*{\qed}
\end{align*}
\renewcommand{\qed}{}
\end{proof}

\begin{cor}
\label{H=E}
Let constraints \eqref{eq:def_qA},  \eqref{eq:qpoisson_A} and the energy conservation $\H=E$ hold. Then we have
\begin{align*}
\hat{J}^{1}_\hbar(x;y;z)
\approx -\frac{\hbar}{i} \frac{a(x,y,z)}{\sqrt{(2p_0)^3}} \frac{\omega}{2\sqrt{2E}} \Q\\
\hat{J}^{2}_\hbar(x;y;z)
\approx +\frac{\hbar}{i} \frac{a(x,y,z)}{\sqrt{(2p_0)^3}} \frac{\omega}{2\sqrt{2E}} \PP \\
\hat{J}^{3}_\hbar(x;y;z)
\approx +\frac{\hbar}{i}  \frac{a^{2}(x,y,z)}{p_0} \frac{\omega}{2\sqrt{2E}}
\tag*{\qed}
\end{align*}
\end{cor}

\begin{rem}
The last Corollary explicitly shows how the quantum theory fundamental law 
$[\pp,\q]=\hbar/i$ spoils the Jacobi identity.
\end{rem}

\section*{Acknowledgements}

The research was in part supported by the Estonian Science Foundation, Grant ETF-6912. The authors are grateful to A. Fialowski for help concerning the Bianchi classification.

\noindent
Department of Mathematics, Tallinn University of Technology,\\
Ehitajate tee 5, 19086 Tallinn, Estonia\\ 
\smallskip
E-mails: eugen.paal@ttu.ee and jvirkepu@staff.ttu.ee

\end{document}